\documentclass[preprint,aps,a4paper]{revtex4-1}
\usepackage{mathrsfs}
\usepackage{graphicx}
\usepackage{multirow}
\usepackage{makecell}
\usepackage{booktabs}
\usepackage{lineno}

\begin{document}
	

     \title{Electronic Origin of High-$T_{C}$ Maximization and Persistence in Trilayer Cuprate Superconductors}

	
    \author{Xiangyu Luo$^{1,3,\dag}$, Hao Chen$^{1,3,\dag}$, Yinghao Li$^{1,3,\dag}$, Qiang Gao$^{1}$, Chaohui Yin$^{1,3}$, Hongtao Yan$^{1,3}$, Taimin Miao$^{1,3}$, Hailan Luo$^{1,3}$, Yingjie Shu$^{1,3}$, Yiwen Chen$^{1,3}$, Chengtian Lin$^{4}$, Shenjin Zhang$^{5}$, Zhimin Wang$^{5}$, Fengfeng Zhang$^{5}$, Feng Yang$^{5}$, Qinjun Peng$^{5}$, Guodong Liu$^{1,3,6}$, Lin Zhao$^{1,3,6}$, Zuyan Xu$^{5}$, Tao Xiang$^{2,3,6,7}$ and X. J. Zhou$^{1,3,6,7,*}$}
		
	\affiliation{
		\\$^{1}$National Lab for Superconductivity, Beijing National laboratory for Condensed Matter Physics, Institute of Physics, Chinese Academy of Sciences, Beijing 100190, China
        \\$^{2}$Beijing National laboratory for Condensed Matter Physics, Institute of Physics, Chinese Academy of Sciences, Beijing 100190, China
		\\$^{3}$University of Chinese Academy of Sciences, Beijing 100049, China
		\\$^{4}$Max Planck Institute for Solid State Research, Heisenbergstrasse 1, D-70569 Stuttgart, Germany
		\\$^{5}$Technical Institute of Physics and Chemistry, Chinese Academy of Sciences, Beijing 100190, China
		\\$^{6}$Songshan Lake Materials Laboratory, Dongguan, Guangdong 523808, China
		\\$^{7}$Beijing Academy of Quantum Information Sciences, Beijing 100193, China
        \\$^{\dag}$These authors contributed equally to this work.
		\\$^{*}$Corresponding author: XJZhou@iphy.ac.cn
	}
	
	\date{\today}
	
	\maketitle
	
	\newpage

{\bf In high temperature cuprate superconductors, it was found that the superconducting transition temperature $T_{c}$ depends on the number of $CuO_{2}$ planes (n) in the structural unit and the maximum $T_{c}$ is realized in the trilayer system (n=3). It was also found that the trilayer superconductors exhibit an unusual phase diagram that $T_{c}$ keeps nearly constant in the overdoped region which is in strong contrast to the $T_{c}$ decrease usually found in other cuprate superconductors. The electronic origin of the $T_{c}$ maximization in the trilayer superconductors and its high $T_{c}$ persistence in the overdoped region remains unclear. By taking high resolution laser-based angle resolved photoemission (ARPES) measurements, here we report our revelation of the microscopic origin of the unusual superconducting properties in the trilayer superconductors. For   the first time we have observed the trilayer splitting in $Bi_{2}Sr_{2}Ca_{2}Cu_{3}O_{10+\delta}$ (Bi2223) superconductor. The observed Fermi surface, band structures, superconducting gap and the selective Bogoliubov band hybridizations can be well described by a three-layer interaction model. Quantitative information of the microscopic processes involving intra- and interlayer hoppings and pairings are extracted. The electronic origin of the maximum $T_{c}$ in Bi2223 and the persistence of the high $T_{c}$ in the overdoped region is revealed. These results provide key insights in understanding high $T_{c}$ superconductivity and pave a way to further enhance $T_{c}$ in the cuprate superconductors.}

      \vspace{3mm}

  Although significant progress has been made in experimental and theoretical studies of high temperature cuprate superconductors, the mechanism of high temperature superconductivity remains a prominent issue in condensed matter physics\cite{PALee2006,BKeimer2015}. In addition to pinning down on the microscopic origin of electron pairing, the challenge also lies in uncovering the key ingredients that dictate high temperature superconductivity. It has been found that the doping level is a key controlling parameter in determining the superconducting transition temperature ($T_{c}$); a maximum $T_{c}$ can usually be observed at the optimal doping (p$\sim$0.16). It has also been found that, even within the same class of cuprate superconductors, the maximum $T_{c}$ depends sensitively on the number of $CuO_{2}$ planes (n) in one structural unit: the maximum $T_{c}$ increases with n from single layer (n=1) to triple layer (n=3), reaches a maximum at n=3 and starts to decrease with further increase of n (Fig. S1a in Supplementary Materials)\cite{AkiraIYO2007,BAScott1994,SChakravarty2004,HEisaki2004,CWChu2015}. For example, in the Bi-based $Bi_{2}Sr_{2}Ca_{n-1}Cu_{n}O_{2n+4+\delta}$ superconductors, the maximum $T_{c}$ increases significantly from 32\,K for n=1 to 91\,K for n=2, to 110\,K for n=3\cite{CWChu2015}. The strong $T_{c}$ dependence on n, particularly the maximum $T_{c}$ enhancement for n=3, indicates that there is another key factor in controlling $T_{c}$ in addition to the doping level. Moreover, the three-layer $Bi_{2}Sr_{2}Ca_{2}Cu_{3}O_{10+\delta}$ (Bi2223) superconductor exhibits an unusual phase diagram in that its $T_{c}$ keeps nearly constant in the optimally and overdoped region (Fig. S1b in Supplementary Materials). This is in a strong contrast to the usual phase diagram where $T_{c}$ decreases with increasing doping in the overdoped region in other one-layer or two-layer cuprate superconductors. Revealing the underlying electronic origin of the $T_{c}$ maximization in trilayer superconductors and its high $T_{c}$ persistence in the overdoped region is significant in understanding superconductivity mechanism and further enhancing $T_{c}$ in cuprate superconductors.

  The Bi2223 superconductor has provided an ideal system for angle-resolved photoemission (ARPES) measurements because of the availability of high quality single crystals and the easiness of cleavage to get clean and flat surface. So far Bi2223 is also the only trilayer superconductor on which extensive ARPES studies have been carried out\cite{RMuller2002,DLFeng2002,TSato2002, HMatsui2003_2,SIdeta2010,SIdeta2010_2,SIdeta2012,SKunisada2017,SIdeta2021}. In Bi2223 with three adjacent $CuO_{2}$ planes in one structural unit (Fig. 1a), it is expected from band structure calculations that three Fermi surface sheets should arise due to interlayer interactions\cite{JACMartinez2014}. However, only one Fermi surface\cite{RMuller2002,DLFeng2002,TSato2002, HMatsui2003_2} or two Fermi surface sheets\cite{SIdeta2010,SIdeta2010_2,SIdeta2012,SKunisada2017,SIdeta2021} have been observed in all the previous ARPES measurements on Bi2223. The absence of three Fermi surface sheets measured in Bi2223 has been attributed to the charge imbalance between the inner and outer $CuO_{2}$ planes and the weak interlayer coupling\cite{SIdeta2010,SIdeta2010_2,SKunisada2017,SIdeta2021}. It was also pointed out that trilayer and higher-multilayer splittings would be increasingly difficult to observe because of the induced pseudogap of the inner layers\cite{SChakravarty2004}.

  In this paper, by taking high resolution laser-based ARPES measurements, we report for the first time the observation of three Fermi surface sheets in Bi2223. The momentum-dependence of the superconducting gap along all the three Fermi surface sheets is determined and the Bogoliubov band hybridization between two specific bands is observed. These observations make it possible to analyse in detail the intralayer and interlayer couplings and pairings. The observed Fermi surface topology, the selective band hybridization and the unusual Fermi surface- and momentum-dependent superconducting gap can be well understood by a three-layer interaction model with a global set of parameters. Our results reveal the microscopic origin of the unusual superconducting properties in the trilayer superconductors.

   The ARPES measurements were carried out on an overdoped Bi2223 sample with a $T_{c}$ of 108.0\,K (Fig. S2 in Supplementary Materials). The high resolution and high statistics data from our laser ARPES system are essential for the new observations we present in the paper (see Methods). Fig. 1 shows the Fermi surface mapping and constant energy contours measured at 18\,K. The corresponding band structures along the momentum cuts from the nodal direction to the antinodal region are presented in Fig. 2. From these results, three main Fermi surface sheets (labelled as $\alpha$, $\beta$ and $\gamma$ in Fig. 1f) and three main bands (marked as $\alpha$, $\beta$ and $\gamma$ in Fig. 2) are clearly observed. In the Fermi surface mapping (Fig. 1b), the spectral weight is mainly confined to the nodal region because of the anisotropic gap opening that is large near the antinodal region. Upon increasing the binding energy, the spectral weight spreads to the antinodal region in the constant energy contours (Fig. 1c-1e) and the full contours of the three main Fermi surface show up clearly. By analyzing the Fermi surface mapping and the constant energy contours in Fig. 1b-1e, combined with the analysis of the band structures in Fig. 2, we have determined quantitatively the three main Fermi surface sheets as shown in Fig. 1f. It is the first time that three main Fermi surface sheets are observed in Bi2223. The $\gamma$ Fermi surface is well separated from the ($\alpha$, $\beta$) sheets. The splitting between the $\alpha$ and $\beta$ sheets increases from the nodal to the antinodal regions, similar to the bilayer splitting observed in $Bi_{2}Sr_{2}CaCu_{2}O_{8+\delta}$ (Bi2212)\cite{DLFeng2001,YDChuang2001,PVBogdanov2001,PAi2019}. The corresponding doping levels of the $\alpha$, $\beta$ and $\gamma$ Fermi surface sheets, determined from their areas, are 0.37, 0.22 and 0.08, respectively.

   Figure 2 shows the band structure evolution with momentum going from the nodal direction to the antinodal region. Three main bands are clearly observed, labelled as $\alpha$, $\beta$ and $\gamma$ and marked by colored arrows in Fig. 2a. Along the nodal direction (first panel in Fig. 2a), the $\gamma$ band is well separated from the ($\alpha$, $\beta$) bands. Although the splitting between the $\alpha$ and $\beta$ bands is a minimum along the nodal direction, careful analyses indicate that the $\alpha$ and $\beta$ band splitting already exists along the nodal direction with a $\Delta k_{F}$=0.011$\pi/a$ (see Fig. S3 in Supplementary Materials). The $\alpha$ and $\beta$ band splitting increases with the momentum moving from the nodal to the antinodal regions. The $\gamma$ band sinks quickly to the high binding energy due to the gap opening, accompanied by strong spectral weight suppression when the momentum cut shifts from the nodal direction to the antinodal region. Near the antinodal region, the $\beta$ band becomes dominant which has much stronger spectral weight than that of the $\alpha$ and $\gamma$ bands. One particularly interesting observation is the selective band hybridization. We find that it is the Bogoliubov back-bending band of the $\beta$ band that hybridizes with the $\gamma$ band. Such a band hybridization becomes observable starting from the momentum Cut 3 (Fig. 2a and 2c), gets stronger with the momentum moving away from the nodal region, reaches the strongest around Cut 6, and then gets weaker with further momentum shifting to the antinodal region. Over the entire momentum space, there is no signature that the $\alpha$ band is involved in the selective Bogoliubov band hybridization.

   In order to determine the superconducting gap along the three main Fermi surface sheets, we show the photoemission spectra (symmetrized energy distribution curves, EDCs) along the three Fermi surface in Fig. 3a-3c (the original EDCs are shown in Fig. S4 in Supplementary Materials). The gap size is obtained from the peak position in the symmetrized EDCs. Since the $\alpha$ band is relatively weak, sitting on top of the strong $\beta$ band, it shows up as shoulders in the symmetrized EDCs as marked by ticks in Fig. 3a. The EDC peaks for the $\beta$ band are strong and obvious along the entire Fermi surface as seen in Fig. 3b. For the $\gamma$ band, prominent EDC peaks are observed along the Fermi surface near the nodal region within the Fermi surface angle $\theta$=45$\sim$26 and then two EDC peaks appear near the antinodal region as marked in Fig. 3c. The superconducting gap along the three Fermi surface sheets, obtained from the symmetrized EDCs in Fig. 3a-3c, is plotted in Fig. 3d. The $\alpha$, $\beta$ and $\gamma$ Fermi surface show quite different maximum gap size of 17, 29 and 62\,meV, respectively, near the antinodal region. Their momentum dependence deviates from a simple $d$-wave form $\Delta=\Delta_{0}\cos(2\theta)$. There appears a discrete jump for the superconducting gap of the $\gamma$ Fermi surface near the antinodal region. As we will explain below, these observations can be naturally understood by considering the interlayer couplings, the interlayer pairings and the selective Bogoliubov band hybridization.

   In principle, for a three-$CuO_{2}$-layer system (Fig. 2e), the Fermi surface, band structures, superconducting gap and band hybridization can be described by the following full Hamiltonian\cite{SIdeta2021}:
   \begin{equation}
    H=\Phi^{\dag}
    \left(
    \begin{array}{cccccc}
    \epsilon_{op}(k)&t_{io}(k)&t_{oo}(k)&\Delta_{op}(k)&\Delta_{io}(k)&\Delta_{oo}(k)\\
    t_{io}(k)&\epsilon_{ip}(k)&t_{io}(k)&\Delta_{io}(k)&\Delta_{ip}(k)&\Delta_{io}(k)\\
    t_{oo}(k)&t_{io}(k)&\epsilon_{op}(k)&\Delta_{oo}(k)&\Delta_{io}(k)&\Delta_{op}(k)\\
    \Delta_{op}(k)&\Delta_{io}(k)&\Delta_{oo}(k)&-\epsilon_{op}(k)&-t_{io}(k)&-t_{oo}(k)\\
    \Delta_{io}(k)&\Delta_{ip}(k)&\Delta_{io}(k)&-t_{io}(k)&-\epsilon_{ip}(k)&-t_{io}(k)\\
    \Delta_{oo}(k)&\Delta_{io}(k)&\Delta_{op}(k)&-t_{oo}(k)&-t_{io}(k)&-\epsilon_{op}(k)\\
    \end{array}
    \right)\Phi
    \end{equation}
     Here the bare bands of the inner plane and the outer planes, $\epsilon_{ip}(k)$ and $\epsilon_{op}(k)$, can be described by the tight-binding model that involves the nearest hopping (t), the second nearest hopping ($t^{'}$) and the third nearest hopping ($t^{''}$) (Fig. 2e). $t_{io}$ represents the interlayer hopping between the inner and outer planes while $t_{oo}$ represents the interlayer hopping between the two outer planes. $\Delta_{ip}$ denotes the intralayer pairing on the inner plane and $\Delta_{op}$ denotes the intralayer pairing on the outer planes. $\Delta_{io}$ represents the interlayer pairing between the inner and outer planes while $\Delta_{oo}$ represents the interlayer pairing between the two outer planes.

   Our first time observation of three Fermi surface sheets (Fig. 1), momentum-dependent band structures and selective Bogoliubov band hybridizations (Fig. 2) and the momentum-dependent superconducting gap along all three Fermi surfaces (Fig. 3) make it possible to extract all these microscopic parameters. In the previous studies of Bi2223 where only two Fermi surface sheets are observed\cite{SIdeta2010_2,SKunisada2017,SIdeta2021}, only the interlayer hopping between the inner and outer planes $t_{io}$ is considered while the hopping between the two outer planes $t_{oo}$ is taken as zero. We started by taking the same approach and tried to fit our data. In this case, a strongly momentum-dependent $t_{io}$ (Fig. S6d) must be assumed in order to get three separated Fermi surface sheets that can well match the measured ones (Fig. S6c). The simulated band structures (Fig. S6a) show significant discrepancies from the measured ones (Fig. S6b). The most notable difference is that, in the simulated band structures (Fig. S6a), the Bogoliubov band hybridization always occurs between the $\alpha$ and $\gamma$ bands while in the measured results (Fig. S6b), the hybridization is actually between the $\beta$ and $\gamma$ bands. In addition, in the simulated band structures (Fig. S6a), the Bogoliubov band hybridization increases dramatically and monotonically from the nodal to the antinodal regions. But in the measured band structures (Fig. S6b), the band hybridization is strong in the intermediate range between the nodal and the antinodal regions and becomes rather weak near the antinodal region. These results indicate that it is impossible to describe our measured results by considering the interlayer hopping between the inner and outer $CuO_{2}$ planes ($t_{io}$) only.

   In order to understand our experimental observations, we find that it is necessary to consider the interlayer hopping between the two outer $CuO_{2}$ planes ($t_{oo}$). Only after this $t_{oo}$ is taken into account can the observed Fermi surface, band structure, superconducting gap and the Bogoliubov band hybridization be well described, as shown in Fig. 2b and 2d. In this case, a strongly anisotropic $t_{oo}(k)=t_{oo0}+t_{oo1}[\cos(k_{x}a)-\cos(k_{y}a)]^{2}/4$ is assumed with $t_{oo0}$=-6\,meV and $t_{oo1}$=-53.3\,meV (red line in Fig. 2f). It is even stronger than the interlayer hopping between the inner and outer planes ($t_{io}$, black line in Fig. 2f). This $t_{oo}$ mainly determines the band splitting between the $\alpha$ and $\beta$ bands, as seen from the diagonalized Hamiltonian (Eq. 8 in Supplementary Materials). Here $t_{oo0}$ is responsible for the slight band splitting along the nodal direction while the band splitting increase from nodal to the antinodal regions originates from the $t_{oo1}$ term. It is interesting to note that $t_{oo}$ has the opposite sign to that of $t_{io}$ (Fig. 2f). After considering $t_{oo}$, the Bogoliubov band hybridization occurs between $\beta$ and $\gamma$ bands (Fig. 2d) which becomes consistent with the measured results (Fig. 2(a,c)). We note that $t_{oo}$ mainly dictates the band splitting between the $\alpha$ and $\beta$ bands while $t_{io}$ plays a dominant role in the Bogoliubov band hybridization.

   The proper considerations of $t_{io}$ and $t_{oo}$ have set a good stage to describe the Fermi surface and band structures of Bi2223 in the normal state. To understand its superconducting state, the intralayer gaps of the inner plane ($\Delta_{ip}$) and the outer planes ($\Delta_{op}$) are usually considered\cite{SKunisada2017,SIdeta2021}. With the discovery of three Fermi surface sheets, particularly the significantly different superconducting gaps between the $\alpha$ and $\beta$ Fermi surface (Fig. 3d), we find that inclusions of $\Delta_{ip}$ and $\Delta_{op}$ alone can not properly describe the observed superconducting gap structure because, in this case, the gap difference between the $\alpha$ and $\beta$ Fermi surface is always small (Fig. S7h in Supplementary Materials) which is much smaller than the observed gap difference near the antinodal region. To overcome such a discrepancy, the interlayer pairing between the two outer planes ($\Delta_{oo}$) has to be considered. After considering $\Delta_{oo}$ which takes a $d$-wave form $\Delta_{oo}$=5$\cos(2\theta)$ (Fig. 2g), the simulated gap structure becomes consistent with the measured one, as seen in Fig. S7p and Fig. S8 in Supplementary Materials. We note that the interlayer pairing between the inner and outer planes ($\Delta_{io}$) is negligible when compared with $\Delta_{oo}$. First, inclusion of $\Delta_{io}$ will reduce the gap size of the $\beta$ band and further reduce the gap difference between the $\alpha$ and $\beta$ bands (Fig. S7(k,l)) which goes against the experimental results. Second, if $\Delta_{io}$ is significant and takes a $d$-wave form, it will enhance the Bogoliubov band hybridization near the antinodal region which is not consistent with the measured result that the hybridization is weak near the antinodal region (Fig. 2(a,c)).

   The above discussions result in the three-layer Hamiltonian (Eq. 7 in Supplementary Materials) which, when the related parameters are properly taken (Fig. 2(f,g)), can satisfactorily describe the observed Fermi surface (Fig. 2b), band structures, superconducting gap and Bogoliubov band hybridizations (Fig. 2d). Further diagonalization of the Hamiltonian (Eq. 8 in Supplementary Materials) indicates that the Bogoliubov band hybridization occurs only between the $\beta$ and $\gamma$ bands but not affected by the $\alpha$ band. This makes it possible to fit the observed $\beta$ and $\gamma$ bands and extract the related parameters quantitatively (see Eqs. 10-12 in Supplementary Materials). Fig. 4b shows the fitted results of the $\beta$ and $\gamma$ bands; detailed fittings of more momentum cuts are shown in Fig. S9 in Supplementary Materials. From the fittings, the anti-crossing gap 2$\Delta_{t\bot}$ defined in Fig. 3g is obtained and plotted in Fig. 4c. It exhibits a nonmonotonic momentum dependence that the band hybridization is strong in the intermediate region between $\theta$=15$\sim$39 but becomes rather weak near the nodal and antinodal regions. Similarly, the extracted $t_{io}$ (Fig. 4d) is also nonmonotonic which is strong in the intermediate region between $\theta$=15$\sim$39 but gets rather weak near the nodal and antinodal regions. The fitting extracted bare superconducting gaps of the $\beta$ and $\gamma$ Fermi surface, $\Delta_{F\beta}$ and $\Delta_{F\gamma}$, are plotted in Fig. 4h. They show obvious difference from the measured superconducting gaps in the region of strong band hybridization. In particular, the fitted gap of the $\gamma$ band ($\Delta_{F\gamma}$) is monotonic and continuous which is different from the observed gap jump due to strong Bogoliubov band hybridization. As seen from the diagonalized Hamiltonian (Eq. 8 in Supplementary Materials), the fitted gap of the $\beta$ band ($\Delta_{F\beta}$) corresponds to $\Delta_{op}+\Delta_{oo}$ while the measured gap of the $\alpha$ band corresponds to $\Delta_{op}-\Delta_{oo}$. Therefore, the interlayer pairing $\Delta_{oo}$ can be directly extracted from the gap difference $2\Delta_{oo}=\Delta_{F\beta}-\Delta_{M\alpha}$ which is plotted in Fig. 4e. The experimentally extracted interlayer hopping $t_{io}$ (Fig. 4d) and interlayer pairing $\Delta_{oo}$ (Fig. 4e) are consistent with those used in the global simulation (Fig. 2f and 2g).

   We find that the interlayer hoppings in the overdoped Bi2223 we measured are quite unusual. First, the interlayer hopping $t_{io}$ exhibits an unusual momentum dependence. Usually, this term is assumed to increase monotonically from the nodal to the antinodal regions\cite{SKunisada2017,SIdeta2021}. But the measured results indicate that it is nonmonotonic and gets strongly suppressed near the antinodal region (Fig. 4d). One possibility is that this may be caused by the intercell hopping. In Bi2223, one cell consists of three $CuO_{2}$ planes and one unit cell contains two cells that are shifted by (a/2, b/2) as seen from the crystal structure in Fig. 1a. Whether the intercell hopping may give rise to such an unusual $t_{io}$ needs further investigations\cite{RSMarkiewicz2005}. The other possibility is that such an unusual $t_{io}$ is intrinsic to the single cell of three $CuO_{2}$ layers which has uneven charge distribution and the inner plane is heavily underdoped with pseudogap formation near the antinodal region. In another Bi-based superconductor $Bi_{2}Sr_{2}CaCu_{2}O_{8+\delta}$ (Bi2212), it has been shown that the single cell can embody the same physical properties as the bulk\cite{YJYu2019}. Further investigations need to be done to pin down on the exact origin of the unusual $t_{io}$ in Bi2223. The second unusual behavior of the interlayer hoppings is that the interlayer hopping between the two outer $CuO_{2}$ planes $t_{oo}$ is significant and even stronger than the interlayer hopping between the inner and outer planes $t_{io}$. Usually, $t_{oo}$ is assumed to be zero in Bi2223 because the distance between the two outer planes is twice that between the inner and outer planes\cite{SKunisada2017,SIdeta2021}. The enhancement of $t_{oo}$ in the overdoped Bi2223 we studied maybe related to the increased doping levels on the outer $CuO_{2}$ planes. It has been shown that, in the multilayer systems, the interlayer hopping depends not only on the distance between the layers but also on the doping levels on each layer\cite{MMori2006}.

   Our present study provides important information on the electronic origin of the maximum $T_{c}$ achieved in trilayer cuprate superconductors. In order to understand the $T_{c}$ maximum in trilayer cuprate superconductors, two main ideas have been proposed. The first is related to the interlayer pair hopping that may enhance superconductivity\cite{SChakravarty2004,KNishiguchi2013}. The second is a composite picture where a high pairing scale is derived from the underdoped planes and a large phase stiffness from the optimally or overdoped ones\cite{SAKivelson2002,EBerg2008,SOkamoto2008}. For superconductors with low superfluid density like the cuprate superconductors, it has been shown that $T_{c}$ is determined not only by the pairing strength but also by the phase coherence\cite{VJEmery1995}. Our results are rather consistent with the composite picture\cite{SAKivelson2002,EBerg2008,SOkamoto2008}. When there is only one $CuO_{2}$ plane in a cell like in $Bi_{2}Sr_{2}CuO_{6+\delta}$ (Bi2201) or there are two equivalent $CuO_{2}$ planes in a cell like in Bi2212, the pairing strength and the phase coherence have to be realized on the same $CuO_{2}$ plane that makes them difficult to be optimized. But in Bi2223, there are two kinds of distinct $CuO_{2}$ planes that makes it possible to separately optimize the pairing strength and phase coherence on different planes. The $\gamma$ Fermi surface originates mainly from the inner plane. It is heavily underdoped (p$\sim$0.08) with a large superconducting gap (Fig. 4h) which may act as the source of strong pairing strength. The $\alpha$ and $\beta$ Fermi surface come mainly from the two outer $CuO_{2}$ planes. These two planes are heavily overdoped (p$\sim$0.30) and may act as a source of strong phase stiffness. It was found that the intensity of the superconducting coherence peak near the antinodal region scales with the superfluid density\cite{DLFeng2000,HDing2001}. Looking at the photoemission spectra along the three Fermi surface sheets in Fig. 3(a-c) and Fig. S4 in Supplementary Materials, we find that the superconducting coherence peaks in the EDCs near the antinodal region along the $\alpha$ and $\gamma$ Fermi surface are weak while they are rather sharp and strong in the EDCs along the $\beta$ Fermi surface. These indicate that the $\beta$ Fermi surface is mainly responsible for providing strong phase coherence in Bi2223. Therefore, in the real space, the composite structure of a heavily underdoped inner $CuO_{2}$ plane (p$\sim$0.08) sandwiched in between two heavily overdoped outer planes (p$\sim$0.30) helps in achieving high temperature superconductivity that no individual $CuO_{2}$ plane can reach. In the reciprocal space, the heavily underdoped $\gamma$ Fermi surface (p$\sim$0.08) with a large superconducting gap and the overdoped $\beta$ Fermi surface (p$\sim$0.22) with strong coherence peaks work together to achieve high $T_{c}$ in Bi2223.

   In a composite system, the interlayer hopping ($t_{\bot}$) plays an important role in realizing superconductivity: the maximum $T_{c}$ can be reached only for an optimized $t_{\bot}$\cite{EBerg2008}. Our extraction of the interlayer hoppings, $t_{io}$ (Fig. 4e) and $t_{oo}$ (Fig. 2f), provides key information to understand high $T_{c}$ in Bi2223. In our overdoped Bi2223, the interlayer hopping $t_{io}$ is slightly larger than that in the optimally-doped Bi2223\cite{SIdeta2021} while the $t_{oo}$ is significantly enhanced when compared with the negligible value in the optimally-doped Bi2223\cite{SIdeta2021}. These results indicate that, among $t_{io}$ and $t_{oo}$, $t_{io}$ is the dominant factor in reaching the maximum $T_{c}$ in Bi2223.

   On the other hand, the interlayer hopping $t_{oo}$ plays a key role in maintaining high $T_{c}$ in the overdoped Bi2223. In the optimally-doped Bi2223, the doping levels of the $\alpha$, $\beta$ and $\gamma$ Fermi surface are 0.23, 0.23 and 0.07 with no observable band splitting between the $\alpha$ and $\beta$ bands\cite{SIdeta2010,SKunisada2017,SIdeta2021}. In our overdoped Bi2223, the doping levels of the $\alpha$, $\beta$ and $\gamma$ Fermi surface are 0.37, 0.22 and 0.08. It becomes immediately clear that, although the average doping increases from 0.18 to 0.22, the extra doping mainly goes to the $\alpha$ Fermi surface, leaving the dopings of the $\beta$ and $\gamma$ Fermi surface nearly unchanged in the overdoped Bi2223 (Fig. 4g). These results indicate that, with the increase of hole doping from the optimally-doped to the overdoped Bi2223, the extra holes mainly goes to the outer $CuO_{2}$ planes. Furthermore, due to the $\alpha$ and $\beta$ band splitting from the interlayer hopping $t_{oo}$, the extra holes are predominantly absorbed into the $\alpha$ Fermi surface. The $\beta$ Fermi surface mainly controls the phase stiffness and its doping changes little in the optimally-doped and overdoped Bi2223 (Fig. 4g). The $\gamma$ Fermi surface mainly controls the pairing strength and its doping levels (Fig. 4g) and the superconducting gap (Fig. 4h) show little change in the optimally-doped and overdoped Bi2223. It is therefore natural to understand that $T_{c}$ exhibits a weak doping dependence in the optimally-doped and overdoped Bi2223 samples (Fig. S1b in Supplementary Materials). In the meantime, the emergence of the interlayer pairing $\Delta_{oo}$ in the overdoped Bi2223 (Fig. 4e) leads to an enhancement of the superconducting gap for the $\beta$ Fermi surface ($\Delta_{M\beta}=\Delta_{op}+\Delta_{oo}$) (Fig. 4h). This in turn will also help in maintaining high $T_{c}$ in the overdoped Bi2223 samples.

   In summary, by taking high resolution laser ARPES measurements, we have observed for the first time the trilayer splitting in Bi2223. The observed Fermi surface, band structures, superconducting gap and the selective Bogoliubov band hybridizations can be well described by a global three layer Hamiltonian. Quantitative information of the microscopic processes involving the intra- and interlayer hoppings and pairings are extracted. The electronic origin of the maximum $T_{c}$ in Bi2223 and the persistence of the high $T_{c}$ in the overdoped region is revealed. These results provide new insights in understanding high $T_{c}$ superconductivity and pave a way to further enhance $T_{c}$ in the cuprate superconductors.\\

  {\bf Methods}

  High quality single crystals of $Bi_{2}Sr_{2}Ca_{2}Cu_{3}O_{10+\delta}$ (Bi2223) were grown by the traveling solvent floating zone method\cite{BLiang2002}. The samples were post-annealed at $550^{o}C$ under high oxygen pressure of $\sim$170 atmospheres for 7 days. The obtained samples are overdoped with a $T_{c}$ of 108.0\,K and a transition width of $\sim$3.0\,K, as seen in Fig. S2 in Supplementary Materials.

  ARPES measurements were carried out using our lab-based laser ARPES systems equipped with the 6.994 eV vacuum-ultra-violet (VUV) laser and a DA30L hemispherical electron energy analyzer\cite{XJZhou2008GDLiu,WTZhang2018XJZhou}. The energy resolution was set at 1\,meV and the angular resolution is $\sim$0.3$^{o}$, corresponding to a momentum resolution of $\sim$0.004 $\AA$$^{-1}$.  All the samples were cleaved $in situ$ at a low temperature and measured in vacuum with a base pressure better than $5\times10^{-11}$ Torr at 18\,K. The Fermi level is referenced by measuring on a clean polycrystalline gold that is electrically connected to the sample.

    \vspace{3mm}

    \noindent {\bf Acknowledgement} This work is supported by the National Natural Science Foundation of China (Grant Nos. 11888101, 11922414 and 11974404), the National Key Research and Development Program of China (Grant Nos. 2021YFA1401800, 2017YFA0302900, 2018YFA0305602 and 2018YFA0704200), the Strategic Priority Research Program (B) of the Chinese Academy of Sciences (Grant No. XDB25000000 and XDB33000000), the Youth Innovation Promotion Association of CAS (Grant No. Y2021006) and the Synergetic Extreme Condition User Facility (SECUF).

    \vspace{3mm}

    \noindent {\bf Author Contributions}\\
    X.J.Z. and X.Y.L. proposed and designed the research. X.Y.L., H.C., C.H.Y., Q.G. and H.T.Y. carried out the ARPES experiments. C.T.L grew the single crystals. H.C., C.H.Y., T.M.M., H.L.L., Y.J.S., Y.W.C., S.J.Z., Z.M.W., F.F.Z., F.Y., Q.J.P., G.D.L., L.Z., Z.Y.X. and X.J.Z. contributed to the development and maintenance of Laser-ARPES and ARToF systems. X.Y.L., Y.H.L., Q.G. and T.X. contributed to theoretical analysis. X.Y.L. and X.J.Z. analyzed the data and wrote the paper. All authors participated in discussions and comments on the paper.

    \vspace{3mm}

    \noindent {\bf\large Additional information}\\
    \noindent{\bf Competing financial interests:} The authors declare no competing financial interests.

    \newpage

    \begin{figure*}[tpb]
    \begin{center}
    	\includegraphics[width=1.0\columnwidth,angle=0]{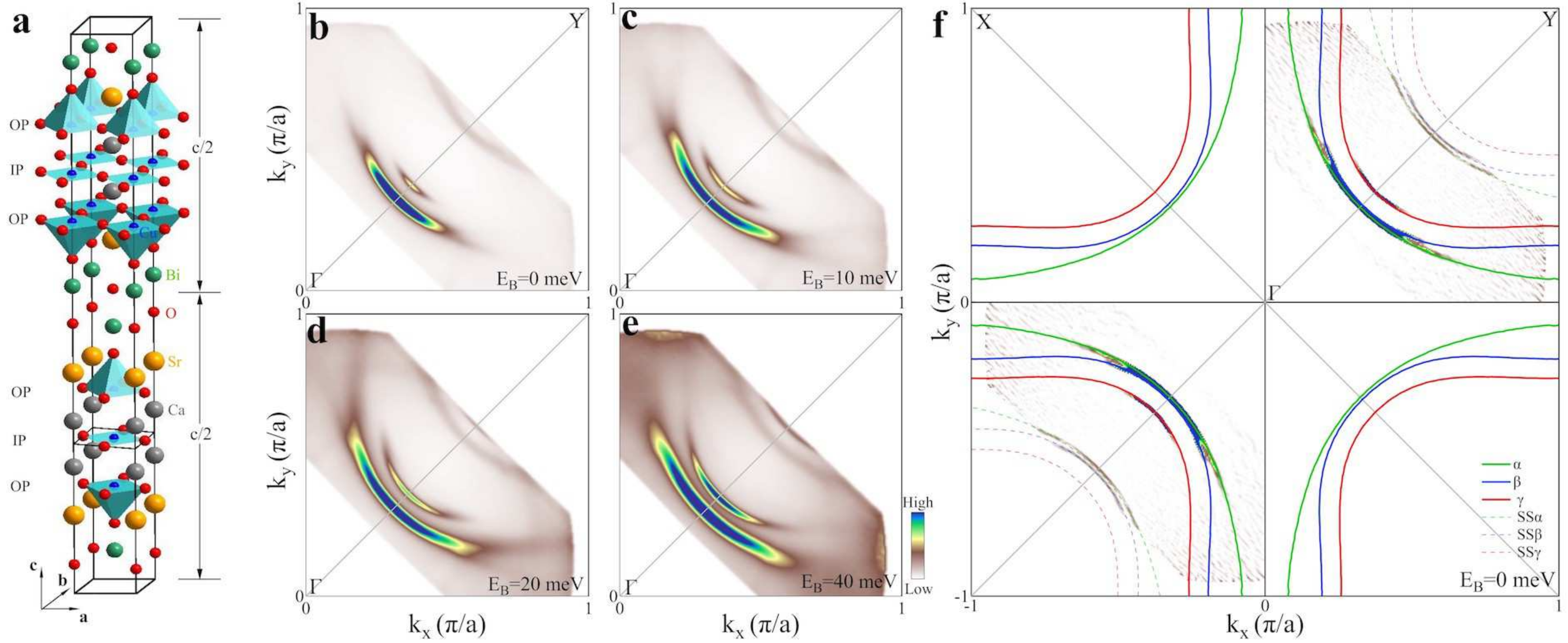}
    \end{center}

    \caption{{\bf Observation of three Fermi surface sheets in Bi2223.} (a) Crystal structure of Bi2223. It consists of two structural units in one unit cell along the c-axis which are displaced with each other by ({\bf a}/2,{\bf b}/2). Each structural unit contains three adjacent $CuO_{2}$ planes with one inner plane (IP) and two outer planes (OP). (b-e) Fermi surface mapping (b) and constant energy contours at binding energies of 10 (c), 20 (d) and 40\,meV (e) measured at 18\,K. (f) Fermi surface of Bi2223 obtained from the analyses of the Fermi surface mappings in (b-e) and related band structures in Fig. 2. The second derivative image of (b) is shown in the first quadrant and the third quadrant. Three Fermi surface sheets are observed labelled as $\alpha$, $\beta$ and $\gamma$. They are fitted by using tight binding model as described in Supplementary Materials and the fitted curves are plotted as solid lines of green ($\alpha$), blue ($\beta$) and red ($\gamma$) colors. The dashed lines represent the first order superstructure Fermi surface ($SS\alpha$, $SS\beta$ and $SS\gamma$) caused by the superstructure modulation in Bi2223 with a wave vector of $Q=(0.247, 0.247)\pi/a$.}

    \end{figure*}

    \begin{figure*}[tbp]
	\begin{center}
		\includegraphics[width=1.0\columnwidth,angle=0]{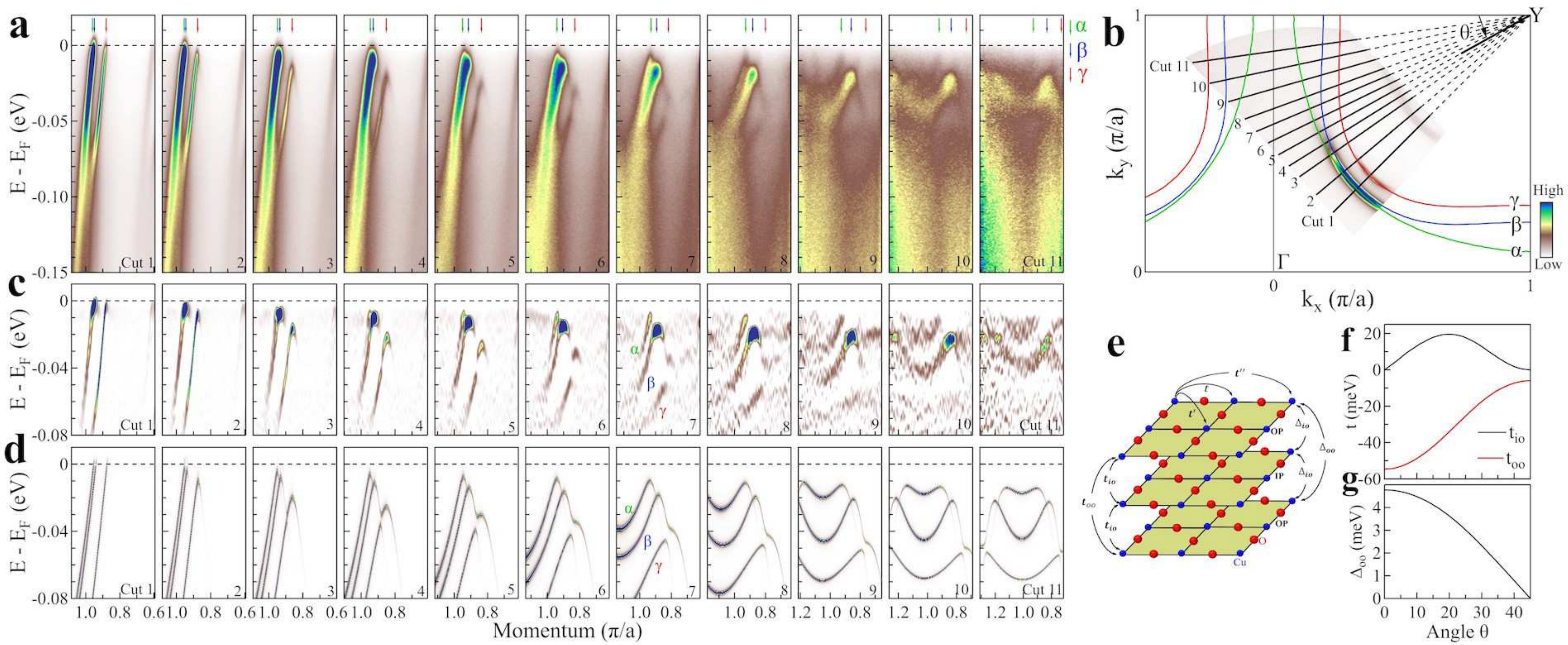}
	\end{center}
	
	 \caption{{\bf Momentum-dependent band structures of Bi2223 measured at 18\,K in the superconducting state and their global simulations.} (a) Band structures measured along different momentum cuts. The location of the momentum cuts is shown in (b) as black lines. These lines all point to the $Y(\pi, \pi)$ point. Three main bands are observed and labelled as $\alpha$, $\beta$ and $\gamma$ that are marked by green, blue and red arrows, respectively. The hybridization of the $\beta$ Bogoliubov back-bending band and the $\gamma$ band is observed clearly for the momentum Cuts 3 to 8. (b) Fermi surface mapping and the simulated three Fermi surface sheets $\alpha$ (green line), $\beta$ (blue line) and $\gamma$ (red line), with the momentum cuts marked. The simulated Fermi surface are in good agreement with the measured Fermi surface. (c) The corresponding EDC second derivative images from (a). (d) Simulated band structures along the same momentum cuts as in (b) by the single set of parameters . Details of the simulation are described in Supplementary Materials. (e) Schematic illustration of the hopping and pairing processes among the three CuO$_{2}$ planes. The $t$, $t^{'}$ and $t^{''}$ represent the in-plane nearest neighbor, the second nearest neighbor and the third nearest neighbor hoppings, respectively. The $t_{io}$ represents the interlayer hopping between the inner CuO$_{2}$ plane (IP) and the outer ones (OP) while the $t_{oo}$ represents the interlayer hopping between the two outer planes. The $\Delta_{io}$ refers to the interlayer pairing between the inner and outer $CuO_{2}$ planes while the $\Delta_{oo}$ refers to the interlayer pairing between the two outer CuO$_{2}$ planes. (f-g) The $t_{io}$ (black line) and $t_{oo}$ (red line) (f) and the interlayer pairing $\Delta_{oo}$ (g) used in the global simulation.}

	\end{figure*}

    \begin{figure*}[tbp]
	\begin{center}
		\includegraphics[width=0.9\columnwidth,angle=0]{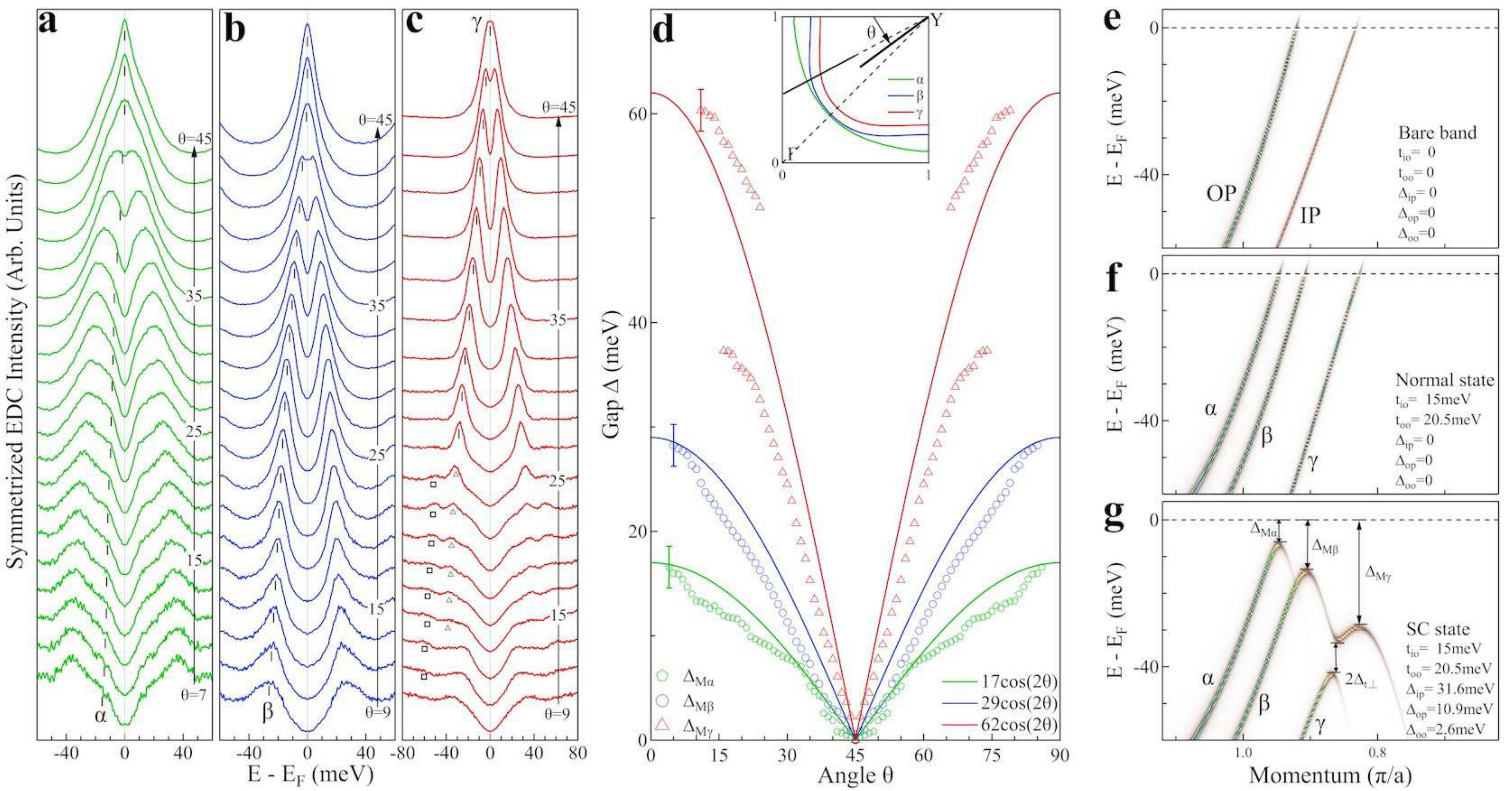}
	\end{center}
	
	\caption{{\bf Photoemission spectra and the superconducting gap of Bi2223 along the three Fermi surface sheets measured at 18\,K.} (a-c) shows symmetrized EDCs along the $\alpha$ (a), $\beta$ (b) and $\gamma$ (c) Fermi surface sheets, respectively. The location of the Fermi momentum is defined by the angle $\theta$, as shown in the inset of (d) where $\theta=0$ corresponds to the $(0, \pi)$ antinodal region while $\theta=45$ corresponds to the nodal region. The symmetrized EDC peaks that correspond to $\alpha$ band are marked by ticks in (a); they are on the shoulders of the symmetrized EDC peaks of the underlying $\beta$ band. The symmetrized EDC peaks that correspond to $\beta$ band are marked by ticks in (b). For the symmetrized EDCs along the $\gamma$ Fermi surface, one main peak is observed near the nodal region with $\theta=45\sim27$ as marked by ticks in (c). Towards the antinodal region with $\theta=25\sim15$, two EDC peaks are observed due to band hybridization, as marked by triangles and squares in (c). (d) Superconducting gaps along the three Fermi surface sheets obtained from the symmetrized EDCs in (a-c). The uncertainty is $\pm$2\,meV as marked on the leftmost points. For comparison with the standard $d$-wave form $\Delta=\Delta_{0}\cos(2\theta)$, three lines are plotted for the $\alpha$, $\beta$ and $\gamma$ Fermi surface with $\Delta_{0}$ of 17(green line), 29 (blue line) and 62\,meV(red line). (e-g) Simulated band structures along a momentum cut in normal and superconducting states. The location of the momentum cut is shown in the inset of (d) as a black line. (e) shows the bare bands of the inner plane (IP) and outer plane (OP). (f) shows three bands that are produced after the introduction of the interlayer couplings of $t_{io}$ and $t_{oo}$. (g) shows the band structures in the superconducting state when the interlayer pairing between the two outer planes, $\Delta_{oo}$, is added on top of $\Delta_{ip}$ and $\Delta_{op}$. The anti-crossing gap 2$\Delta_{t\bot}$ is also defined. Note that the Fermi momentum of the three bands keeps fixed from the normal state in (f) to the superconducting state in (g).
}
	
	\end{figure*}

    \begin{figure*}[tbp]
	\begin{center}
		\includegraphics[width=0.9\columnwidth,angle=0]{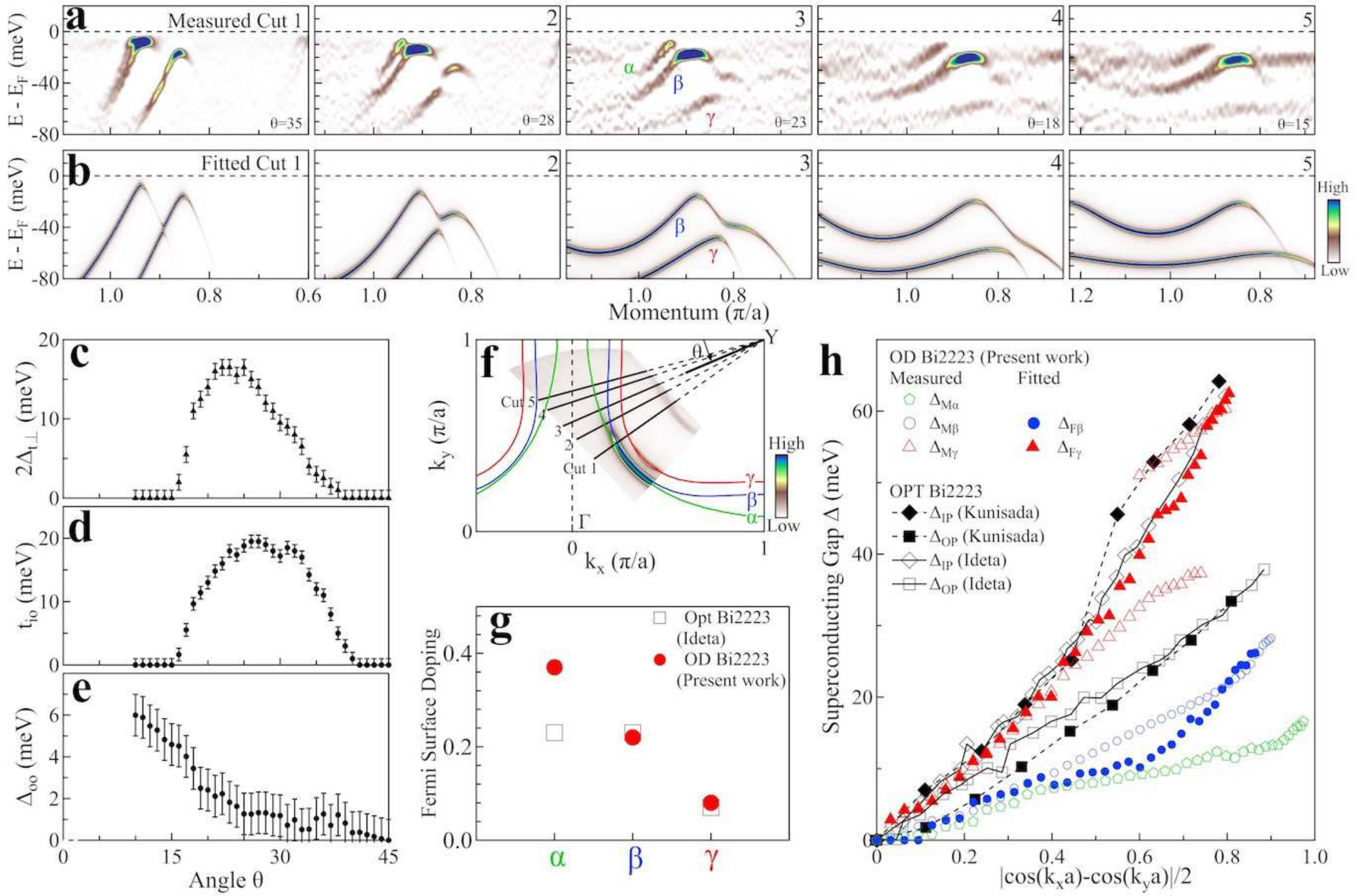}
	\end{center}
	
	\caption{{\bf Determination of the interlayer hopping, interlayer pairing and band hybridization parameters in Bi2223.} (a) Band structures measured along five typical momentum cuts. The location of the momentum cuts is marked by black lines in (f). These are EDC second derivative images which can show the band hybridization between $\beta$ and $\gamma$ bands more clearly. (b) Fitted band structures along the same five momentum cuts as in (a) by considering the band hybridization of the $\beta$ and $\gamma$ bands. The fitting details are described in Supplementary Materials. (c-d) The anti-crossing gap $2\Delta_{t\perp}$(c) and the interlayer hopping $t_{io}$ (d) as a function of angle $\theta$ extracted from fitting the band hybridizations in (a), as shown in (b) and Fig. S9 in Supplementary Materials. (e) The interlayer pairing $\Delta_{oo}$ obtained from the superconducting gaps along the $\alpha$ and $\beta$ Fermi surface sheets shown in (h). (f) Fermi surface of Bi2223 and the location of the momentum cuts. (g) The doping levels corresponding to the Fermi surface sheets in our overdoped Bi2223 (red solid circles) and in the optimally-doped Bi2223 (black empty squares)\cite{SIdeta2010}. (h) The superconducting gaps as a function of $|cos(k_{x}a)-cos(k_{y}a)|/2$ along the three Fermi surface in our overdoped Bi2223 shown in Fig. 3d. The fitted superconducting gaps along the $\beta$ ($\Delta_{F\beta}$, solid blue circles) and $\gamma$ ($\Delta_{F\gamma}$, solid red triangles) Fermi surface sheets extracted from the band structure fitting in (b) and in Fig. S9 in Supplementary Materials are included. For comparison, the superconducting gaps measured in the optimally-doped Bi2223\cite{SKunisada2017,SIdeta2021} are also plotted.
}

	\end{figure*}

\end{document}